\documentclass[sigconf,nonacm]{acmart}
\usepackage{amsmath,booktabs,graphicx,float}
\usepackage{balance}
\setcopyright{none}
\renewcommand{\footnotetextcopyrightpermission}[1]{}
\acmDOI{}
\acmISBN{}
\makeatletter
\let\@ACM@checkaffil\relax
\makeatother
\makeatletter
\renewcommand{\@mkauthors@iii}{%
  \global\setbox\mktitle@bx=\vbox{%
    \unvbox\mktitle@bx
    \centering
    {\large\def\and{\quad}\authors\par}
    {\small AMAP, Alibaba Group\par}
    {\footnotesize
      \texttt{\{chuhuan.zy,lw543345,zhangxuewen.zxw,yanzhuang.yz,pengjie.wpj\}@alibaba-inc.com}\par}
    \smallskip}}
\makeatother
\hypersetup{
  pdfauthor={Yi Zou, Wei Luo, Xuewen Zhang, Zhuang Yan, Pengjie Wang},
  pdfkeywords={Cross-domain POI recommendation, implicit intent inference, mobility behavior, large language models, reinforcement learning}
}
\usepackage{colortbl}
\newcommand{\method}{MILES-Rec}
\newcommand{\benchmark}{MILES-Bench}
\begin{document}
\title{MILES-Rec: Uncovering Implicit Intents from Mobility Behaviors for Cross-Domain POI Recommendation}
\author{Yi Zou}
\affiliation{\institution{AMAP, Alibaba Group}}
\email{chuhuan.zy@alibaba-inc.com}

\author{Wei Luo}
\affiliation{\institution{AMAP, Alibaba Group}}
\email{lw543345@alibaba-inc.com}

\author{Xuewen Zhang}
\affiliation{\institution{AMAP, Alibaba Group}}
\email{zhangxuewen.zxw@alibaba-inc.com}

\author{Zhuang Yan}
\affiliation{\institution{AMAP, Alibaba Group}}
\email{yanzhuang.yz@alibaba-inc.com}

\author{Pengjie Wang}
\affiliation{\institution{AMAP, Alibaba Group}}
\email{pengjie.wpj@alibaba-inc.com}

\renewcommand{\shortauthors}{Zou et al.}
\begin{abstract}
Map applications bridge everyday mobility with local services, creating opportunities to recommend places by uncovering users' implicit intents.
However, utility-driven map interactions provide only indirect and incomplete evidence of consumption demands, making it challenging to infer personalized service needs from observed behaviors.
We propose \method, a cross-domain POI recommendation framework that infers latent service intents from heterogeneous map behaviors and expresses them as prioritized natural language queries.
It first learns from progressive reasoning demonstrations that synthesize user profiles and map histories for situation understanding, intent expansion, and query prioritization.
We then introduce intent-aligned group relative policy optimization, using subsequent user interactions to reward query relevance and ranked service coverage while penalizing redundancy and excessive response length.
A two-level credit assignment mechanism integrates shared list-level advantages with query-specific advantages, coordinating individual query quality with the complementary coverage of the complete query list.
We further construct \benchmark\ from real user interaction logs on a large-scale mapping platform, supporting training and evaluation across diverse services under rich and sparse behavioral histories.
Extensive experiments against general-purpose LLMs and specialized recommendation baselines demonstrate superior POI retrieval utility and strong service and semantic alignment, with consistent gains from behavioral policy optimization across four backbone scales.
\end{abstract}

\maketitle
\hypersetup{pdfauthor={Yi Zou, Wei Luo, Xuewen Zhang, Zhuang Yan, Pengjie Wang}}
\flushbottom
\section{Introduction}

\begin{figure}[t]
    \centering
    \includegraphics[width=\columnwidth]{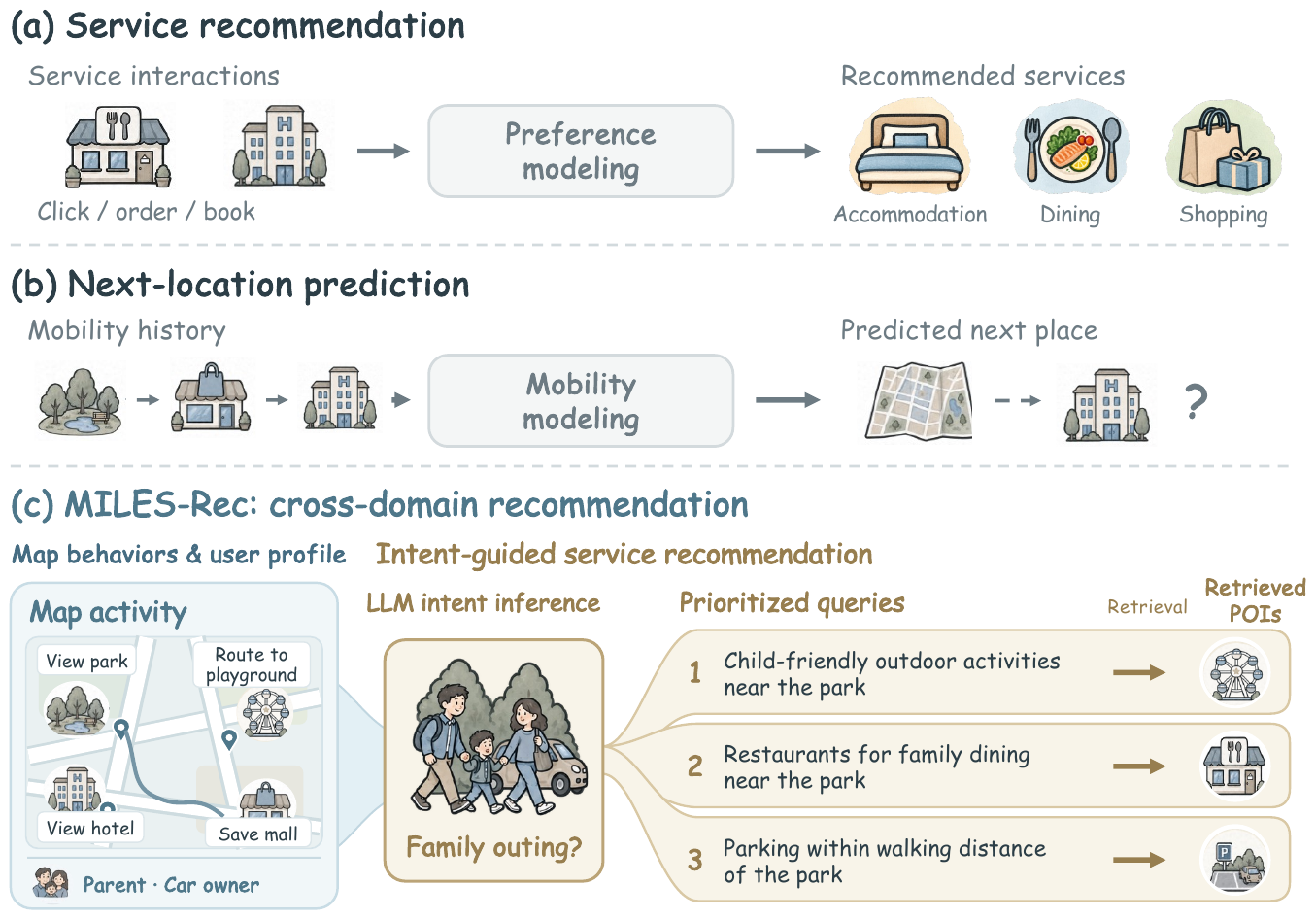}
    \caption{(a) Service recommendation; (b) next-location prediction;
    (c) \method\ infers service intents from map behaviors
    and retrieves POIs using prioritized queries.}
    \label{fig:teaser}
\end{figure}

Local-service recommendation helps users discover places for dining, shopping, leisure,
and other everyday activities. Modern map applications naturally encompass two distinct
yet coupled functional domains: the spatial mobility domain, which records
high-frequency utility actions such as destination searches, route requests, and
navigation, and the local-service domain, which captures service exploration and
consumption across diverse commercial categories. Because users rely on map platforms
primarily as mobility tools, explicit service transactions are often sparse, whereas
mobility trajectories carry rich contextual cues about companion lifestyle needs.
Conventional next-POI prediction focuses on forecasting a user's immediate spatial
destination within the mobility domain~\cite{getnext}, yet it largely overlooks the
derivative service demands triggered by these trips. For instance, planning a route to a
family park often entails subsequent target-domain needs for child-friendly dining or
nearby parking. Bridging these two domains requires jointly interpreting heterogeneous
map histories alongside user profiles to uncover latent consumption intents. This
motivates cross-domain POI recommendation from mobility behaviors to local services,
which leverages abundant mobility evidence to proactively recommend relevant everyday
services beyond explicitly stated user queries, as illustrated in Figure~\ref{fig:teaser}.

Large language models combine broad world knowledge with reasoning capabilities,
offering a strong semantic foundation for linking observed mobility behaviors to
implicit service needs. Expressing these latent connections in natural language provides
an interpretable bridge between source-domain mobility evidence and target-domain
service preferences. Recent studies have explored explicit intent modeling and
query-based recommendation from various perspectives. Specifically, PIGEON~\cite{pigeon}
predicts open-set living needs to retrieve local services via embeddings, AIR~\cite{air}
extracts atomic intents to transfer preferences across content and e-commerce domains,
and EAGER~\cite{eager} aligns supervised query generation with single-domain click
preferences through reinforcement learning. Although these works demonstrate the promise
of explicit intent representations and behavioral optimization, they primarily generate
isolated or unranked intents from homogeneous item interactions. Consequently, a
critical challenge remains in how to infer a prioritized set of complementary service
intents from noisy, heterogeneous map logs and jointly optimize their list-wise
target-domain utility through subsequent multi-service feedback.

Addressing this cross-domain task entails three fundamental challenges. First,
contextualized multi-intent reasoning requires jointly interpreting heterogeneous
mobility actions and user profiles in context. The model must distinguish user-specific
service demands from generic commonsense associations, derive complementary expansion
intents alongside primary needs, and establish their relative priorities. Second,
behavioral supervision across domains is inherently indirect because interaction logs
record only external actions rather than underlying motivations. Since multiple
plausible query sets can correspond to the same mobility activity, merely imitating
static reference queries via supervised learning fails to directly optimize alignment
with subsequent target-domain interactions. Third, ranked query generation involves a
trade-off between individual query relevance and overall list coverage. Optimizing
solely for relevance often yields repetitive queries targeting a single dominant
service, whereas pursuing broad coverage risks introducing weak queries. Because a
single response-level reward conflates these factors across the entire output,
fine-grained credit assignment is required to disentangle individual query quality from
the collective utility of the complete list.

To overcome these challenges, we propose MILES-Rec, a two-stage framework that bridges
the mobility domain with local-service discovery through prioritized natural-language
queries. To support contextualized multi-intent reasoning, we construct progressive
reasoning demonstrations that guide the model through situation understanding, intent
expansion, and query prioritization via supervised fine-tuning. To learn effectively
from indirect behavioral supervision, we introduce intent-aligned Group Relative Policy
Optimization, refining the supervised policy with feedback from subsequent target-domain
interactions. The behavioral reward jointly measures query-to-service relevance and
one-to-one ranked service coverage while penalizing redundancy and excessive length.
Furthermore, we design a two-level credit assignment mechanism to coordinate individual
relevance with list coverage. A global list-level advantage guides the entire response
including reasoning and structure, whereas a local query-specific advantage is
selectively applied to the corresponding query-text tokens. At inference time, MILES-Rec
generates prioritized service queries from user profiles and map histories to retrieve
relevant local-service POIs.

We further construct MILES-Bench from real-world user profiles and heterogeneous
interaction logs on a large-scale mapping platform. It provides 19,006 verified
reasoning demonstrations for supervised fine-tuning, 3,000 interaction-grounded samples
for reinforcement learning, and two user-disjoint test sets covering rich and sparse
histories across ten cities. To assess both intermediate intent reasoning and downstream
retrieval, we establish a unified evaluation protocol spanning POI Retrieval Utility
over a two-million-POI pool, Service Intent Alignment, and Semantic Demand Alignment.
Together, these dimensions evaluate retrieved POI utility alongside the category
relevance, constraint compatibility, and service coverage of the generated
natural-language intent queries.

Our main contributions are summarized as follows:
\begin{itemize}
    \item We formulate cross-domain POI recommendation from the spatial mobility domain
    to the local-service domain as an intent-based reasoning and prioritized query
    generation task over heterogeneous map histories and profiles.
    \item We propose MILES-Rec, which integrates progressive reasoning supervision with
    intent-aligned reinforcement learning and employs a two-level credit assignment
    mechanism to jointly optimize individual query relevance and list-wise service
    coverage while suppressing redundancy.
    \item We construct MILES-Bench from real-world map logs covering both rich and
    sparse histories, paired with a three-dimensional evaluation protocol linking POI
    retrieval utility with service and semantic alignment.
    \item Extensive experiments across four backbone scales demonstrate consistent gains
    over supervised fine-tuning and strong improvements over state-of-the-art baselines,
    supported by comprehensive ablations, human evaluations, and retriever-transfer
    analyses.
\end{itemize}

\section{Related Work}
\label{sec:related_work}

\subsection{LLM-based Recommendation}
Large language models provide a strong semantic foundation for modeling user behaviors
and generating personalized recommendations. Early works unify recommendation tasks
within text-to-text frameworks as in P5~\cite{p5} or adapt language models via
instruction tuning as in TALLRec~\cite{tallrec}, while LLaRA~\cite{llara} and
ReLLa~\cite{rella} further incorporate collaborative embeddings and retrieved
historical interactions to mitigate long-sequence noise. Moving beyond supervised
adaptation, recent studies optimize generation with downstream feedback via
reinforcement learning, including Rec-R1 with fixed-recommender rewards~\cite{rec_r1},
EAGER with click-preference alignment~\cite{eager}, and ReList with multi-objective query
optimization~\cite{relist}. However, these reinforcement learning methods typically
rely on coarse response-level rewards over homogeneous item clicks. In contrast,
MILES-Rec infers implicit service intents from heterogeneous map logs and user profiles,
introducing a two-level credit assignment mechanism within group relative policy
optimization (GRPO) to jointly optimize individual query relevance and list-wise
service coverage without redundant intent generation.

\subsection{Location-based Recommendation}
Location-based recommender systems leverage user preferences and spatiotemporal contexts
to suggest relevant places and services. One mainstream direction focuses on next-POI
prediction within the mobility domain via sequential trajectory modeling~\cite{getnext}
or language-model-based spatiotemporal reasoning~\cite{aware,ros,coast}. Because trajectory
forecasting alone cannot capture companion consumption demands, a complementary line of
research explicitly models living needs for local-service discovery, evolving from
predefined service categories in NEON~\cite{neon} to open-set need descriptions in
PIGEON~\cite{pigeon}, hierarchical agentic service reasoning~\cite{local_life_agent}, and
spatiotemporally constrained intent sequences~\cite{gplan}. Although these studies advance
explicit demand modeling, they primarily predict isolated needs or next spatial
destinations from homogeneous check-in sequences. Our work instead jointly interprets
heterogeneous mobility interactions alongside user profiles, uncovering both primary
needs and complementary expansion intents as a prioritized query list for personalized
multi-service POI discovery.

\subsection{Cross-Domain Recommendation}
Cross-domain recommendation transfers behavioral knowledge across domains to alleviate data sparsity and improve preference modeling. Recent approaches leverage language models to capture semantic relationships beyond item-identifier overlaps, such as combining unified semantic representations with hierarchical profiling in LLM4CDSR~\cite{liu2025bridge} or generating synthetic cross-domain interactions in LLMCDSR~\cite{xin2025llmcdsr}. Closer to our formulation, AIR extracts atomic intents from content browsing to support e-commerce recommendation~\cite{air}. While these methods transfer preferences across item-consumption domains via latent embeddings or unranked atomic tags, they do not address the functional gap between tool-oriented mobility behaviors and commercial service consumption. MILES-Rec bridges the spatial mobility domain with the local-service domain by reasoning over map histories and user profiles to formulate prioritized queries for hybrid spatial retrieval.

\section{Problem Formulation}
\label{sec:problem}

We study cross-domain POI recommendation from the spatial mobility domain to the
local-service domain. The source context is dominated by
high-frequency mobility interactions alongside user profiles, while the target
space encompasses local-service POIs across commercial categories. Because utility-oriented
mobility actions and service POIs lack direct item-level overlap, prioritized natural-language
queries serve as an explicit semantic bridge connecting behavioral observations to
personalized service retrieval.

For each user $u$, we group the observed profile attributes as
\begin{equation}
\mathcal{P}_u =
(\mathbf{p}_u^{\text{basic}}, \mathbf{p}_u^{\text{life}}, \mathbf{p}_u^{\text{vehicle}}).
\label{eq:user_profile}
\end{equation}
Here, $\mathbf{p}_u^{\text{basic}}$ contains age group, occupation, and city of residence;
$\mathbf{p}_u^{\text{life}}$ captures marital status, parenthood, life stage, children's
age groups, and pet ownership; and $\mathbf{p}_u^{\text{vehicle}}$ specifies car ownership,
brand, and model series. Each group includes observed attributes.

The chronological interaction history observed before the prediction time $t_0$ is
$\mathcal{H}_u = [e_{u,1}, \dots, e_{u,n_u}]$, where $n_u \ge 0$ is the number
of historical events and may be zero for cold-start users. Each event
$e_{u,i} = (t_{u,i}, b_{u,i}, x_{u,i})$ records a timestamp, an action type, and
textual content. Historical actions are dominated by source-domain mobility behaviors
such as destination searches, detail views, saves, route planning, and navigation
alongside sparse historical service bookings, while the textual content $x_{u,i}$ includes
search queries and place descriptions. Given the spatiotemporal context
$\mathbf{z}_u = (t_0, \ell_u)$ with available city context $\ell_u$, the complete input state is
defined as
\begin{equation}
\begin{aligned}
s_u &= (\mathcal{P}_u, \mathcal{H}_u, \mathbf{z}_u),
\quad \mathbf{z}_u = (t_0, \ell_u),\\
t_{u,i} &< t_0, \quad \forall i \in \{1, \dots, n_u\}.
\end{aligned}
\label{eq:user_context}
\end{equation}

Given $s_u$, a policy model $f_\theta$ reasons over the user's profile,
mobility trajectory, and context to infer implicit service needs, expressing them as an
ordered query list:
\begin{equation}
\mathcal{Q}_u = f_\theta(s_u) = [q_{u,1}, \dots, q_{u,K_u}],
\label{eq:query_generation}
\end{equation}
where $K_u$ is the number of generated queries. Each query expresses a
distinct service intent and contextual constraints in natural
language, and its position indicates its priority. The generator aims to identify primary intents grounded in recent
mobility actions and uncover complementary expansion intents for unseen services.

Given a shared pool of local-service POIs $\mathcal{V}$, ranking candidates
by relevance to $\mathcal{Q}_u$ yields the final recommendation:
\begin{equation}
\hat{\mathcal{Y}}_u
= \text{Rank}_{B_u}(\mathcal{V}; \mathcal{Q}_u, \mathbf{z}_u)
= [v_{u,1}, \dots, v_{u,B_u}],
\label{eq:poi_recommendation}
\end{equation}
where $B_u$ is the number of returned POIs and $v_{u,j} \in \mathcal{V}$. Target-domain
interactions observed at $t \ge t_0$ are excluded from $s_u$ and provide
behavioral feedback for policy optimization and evaluation.

\section{Method}
\subsection{Overview}
\label{sec:method_overview}

\begin{figure*}[!t]
 \centering
 \includegraphics[width=\textwidth]{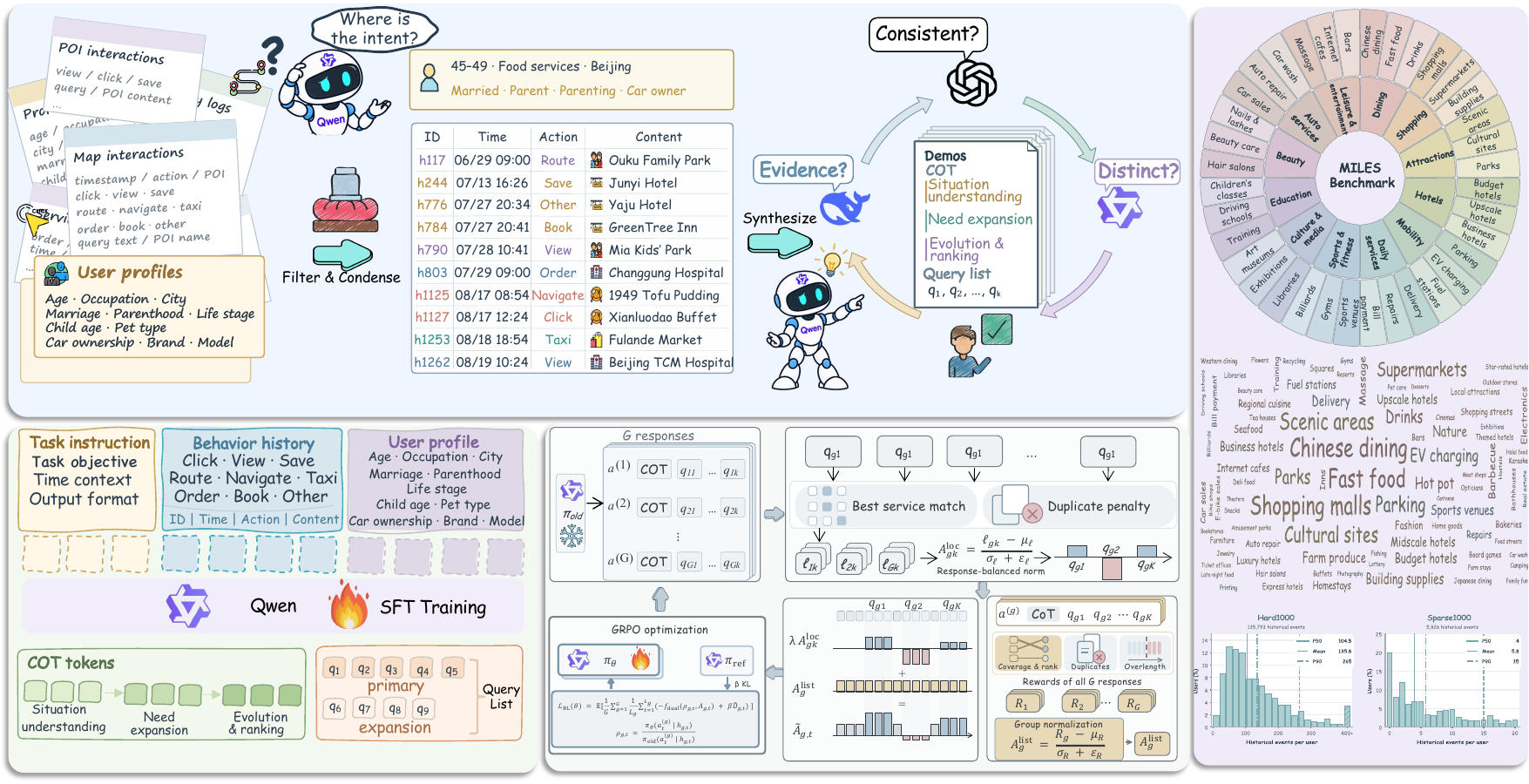}
\caption{Overview of \method: reasoning-supervised SFT and intent-aligned GRPO. Right: \benchmark\ service taxonomy and history distributions.}
 \label{fig:overview}
\end{figure*}

Figure~\ref{fig:overview} illustrates the architecture of \method, which learns
cross-domain POI recommendation from the spatial mobility domain to the local-service
domain via a two-stage paradigm. In Section~\ref{sec:data}, we construct progressive
reasoning demonstrations covering situation understanding, intent expansion, and query
prioritization to initialize the policy through supervised fine-tuning. In
Section~\ref{sec:rl}, we refine the policy with subsequent target-domain interactions
via Group Relative Policy Optimization, introducing two-level credit assignment to
jointly optimize individual query relevance and list-wise service coverage.

\subsection{Progressive Reasoning Supervision}
\label{sec:data}
Directly generating target-domain service queries from mobility logs and user profiles
is challenging because mobility actions primarily reflect movement rather than explicit
consumption requests. We therefore construct progressive chain-of-thought
demonstrations~\cite{cot} that decompose intent inference into three stages before
producing prioritized queries for personalized POI retrieval.

\paragraph{Three-Stage Reasoning Structure.}
\emph{Situation understanding} jointly interprets profile attributes, chronological
interactions, and prediction context to infer the user's underlying activity or circumstances,
citing supporting events. \emph{Intent expansion} identifies primary service needs
grounded in observed behaviors and derives complementary, alternative, or subsequent
intents supported by the context, emphasizing distinct service purposes rather than
synonymous queries. \emph{Query prioritization} considers evidence strength, recency,
and recurring behavioral patterns to organize up to ten concise queries. Primary
queries precede related extensions, and the final list retains a \texttt{primary} or
\texttt{expansion} label for each generated service query.

\paragraph{Demonstration Synthesis and Verification.}
We organize user-level platform records into a structured context $\widetilde{s}_u$
containing available profile attributes, prediction context, and chronological
interactions. Each event retains its identifier, timestamp, action type, and original
content. We remove missing-value placeholders and duplicated contextual fields while
preserving repeated interactions and their temporal order. This reduces input redundancy
and supports event-based evidence citation. Given $\widetilde{s}_u$, Qwen-235B generates
the three-stage reasoning and prioritized queries in a single response.

Building on model-generated supervision and AI feedback~\cite{self_instruct,ultrafeedback},
we refine these demonstrations using GPT-5.1 and DeepSeek-Pro feedback alongside human
review and revision. The review examines evidence fidelity, reasoning consistency, and
intent distinctness, guiding corrections to unsupported attributes, inconsistent
inferences, and redundant queries. The resulting dataset contains 19,006 user-level
demonstrations, each represented as
$a_u^\star=(r_u^\star,\mathcal Q_u^\star)$,
where $r_u^\star$ contains relevant profile facts, cited events, and reasoning,
and $\mathcal Q_u^\star$ contains ordered service queries and their priority labels.

\paragraph{Supervised Fine-Tuning Objective.}
We fine-tune the policy $\pi_\theta$ on complete target responses using token-level negative log-likelihood:
\begin{equation}
\mathcal L_{\mathrm{SFT}}(\theta)
=
-\frac{1}{Z}
\sum_{u\in\mathcal U_{\mathrm{SFT}}}
\sum_{t=1}^{L_u}
\log \pi_\theta
\left(
a_{u,t}^\star
\mid
\widetilde{s}_u,a_{u,<t}^\star
\right),
\label{eq:sft_objective}
\end{equation}
where $\mathcal U_{\mathrm{SFT}}$ denotes the training users,
$L_u$ is the target response length, and
$Z=\sum_{u\in\mathcal U_{\mathrm{SFT}}}L_u$.
The loss covers reasoning, queries, priority labels, and output structure, excluding all input tokens.

This supervision initializes a policy that connects mobility evidence with prioritized
service intents. Subsequent RL explores alternative intent combinations and query
priorities, using later user interactions to refine query relevance and service coverage.

\subsection{Intent-Aligned GRPO}
\label{sec:rl}

We use subsequent target-domain interactions to align generated queries with observed
service needs. The behavioral reward combines individual query relevance with ranked
service coverage, balancing intent quality and the breadth of the query list.
Two-level credit assignment combines shared response-level guidance with additional
local feedback for individual service queries.

Starting from the SFT initialization, we follow GRPO~\cite{deepseekmath} and sample $G$ responses
$a^{(g)}=(r^{(g)},\mathcal Q^{(g)})$
from the rollout policy
$\pi_{\mathrm{old}}(\cdot\mid\widetilde{s}_u)$.
Each response contains a reasoning trace and an ordered query list
$\mathcal Q^{(g)}=[q_{g1},\ldots,q_{gK_g}]$.
Subsequent interactions are used solely to compute rewards and are excluded from the policy input.
\paragraph{Behavioral Reward Construction.}
To connect open-ended intents with observable target-domain interactions,
we adopt 263 fine-grained service categories organized into 12 broad groups
covering dining, shopping, hotels, leisure and entertainment, transportation, automotive
services, and other local services. We group POIs from subsequent interactions by service category into
$\mathcal T=\{T_j\}_{j=1}^{M}$,
with category descriptions $\mathcal C_j$ and observed POI texts $\mathcal D_j$ retained
for each group. These groups provide common coverage targets for distinct POIs
offering the same target-domain service.

We fine-tune BGE-reranker-v2-m3~\cite{bge_reranker} on Qwen-235B-annotated
query--category pairs to assess service alignment across three relations, namely match,
compatible, and mismatch. A pretrained Qwen3-Embedding-8B~\cite{qwen3_embedding}
encoder $E$ additionally measures query--POI semantic similarity. Both models remain
frozen during reinforcement learning; annotation and matcher-training details appear
in Appendix~\ref{sec:grpo_details}. For query $q_{gk}$ and service group $T_j$,
we compute
\begin{equation}
\begin{aligned}
m_{gkj}
&=
\max_{c\in\mathcal C_j}
\left[
P_{\mathrm{match}}(q_{gk},c)
+\alpha P_{\mathrm{compat}}(q_{gk},c)
\right],\\
v_{gkj}
&=
\left[
\max_{d\in\mathcal D_j}
\cos\left(E(q_{gk}),E(d)\right)
\right]_{[0,1]},\\
s_{gkj}
&=m_{gkj}v_{gkj}.
\end{aligned}
\label{eq:matching}
\end{equation}
Here, $P_{\mathrm{match}}$ and $P_{\mathrm{compat}}$ denote the predicted probabilities
of the match and compatible relations, $\alpha=0.5$, and $[\cdot]_{[0,1]}$ clips values
to $[0,1]$. The product $m_{gkj}v_{gkj}$ favors queries supported by both service
agreement and semantic similarity to observed POI texts.

We reward complementary coverage through maximum-weight one-to-one matching
between queries and service groups. Let $\delta_{gk}$ indicate whether
$q_{gk}$ repeats an earlier query, and let $\mathfrak M_g$ contain all feasible partial
matchings. We define ranked coverage as
\begin{equation}
C_g
=
\frac{1}{M}
\max_{\mathcal M\in\mathfrak M_g}
\sum_{(k,j)\in\mathcal M}
\frac{(1-\delta_{gk})s_{gkj}}{\log_2(k+1)}.
\label{eq:list_coverage}
\end{equation}
Each query matches at most one service group, and each group contributes at most once. The rank discount encourages placing well-supported intents earlier, while the one-to-one constraint prevents repeatedly rewarding the same target service.

For valid responses, we also penalize redundancy and excessive response length,
yielding the following list-level reward:
\begin{equation}
R_g
=
\left[
C_g
-\lambda_{\mathrm{dup}}N_g^{\mathrm{dup}}
-L_{\mathrm{len}}(a^{(g)})
\right]_+,
\label{eq:list_reward}
\end{equation}
where $N_g^{\mathrm{dup}}=\sum_k\delta_{gk}$ and $[x]_+=\max(0,x)$.
Repeated queries incur an explicit penalty, while
$L_{\mathrm{len}}$ applies a bounded linear penalty beyond a free response-length
budget, counting both reasoning and queries. Invalid or truncated responses receive
zero reward. Appendix~\ref{sec:grpo_details} specifies the validity criteria and penalty parameters.

\paragraph{Two-Level Credit Assignment.}
A list-level reward evaluates the complete output but does not distinguish the
relevance of individual queries within it~\cite{rank_grpo}. We therefore derive a
local relevance score for each query from the same service-matching matrix:
\begin{equation}
\ell_{gk}
=
\begin{cases}
\max_j s_{gkj}, & \delta_{gk}=0,\\
-\lambda_{\mathrm{dup}}, & \delta_{gk}=1.
\end{cases}
\label{eq:local_score}
\end{equation}
This score measures a query's best service agreement without one-to-one matching or
rank discounting, complementing list-level coverage with an assessment of individual
query relevance.

We normalize each signal within the user's response group:
\begin{equation}
A_g^{\mathrm{list}}=\frac{R_g-\mu_R}{\sigma_R+\varepsilon_R},\qquad
A_{gk}^{\mathrm{loc}}=\frac{\ell_{gk}-\mu_\ell}{\sigma_\ell+\varepsilon_\ell}.
\label{eq:advantages}
\end{equation}
Here, $\mu_R$ and $\sigma_R$ are computed over all $G$ response rewards.
For local statistics, let $\mathcal G_{\mathrm v}$ index valid responses. We compute the weighted mean $\mu_\ell$ and standard deviation $\sigma_\ell$ using weights
$w_{gk}=1/(|\mathcal G_{\mathrm v}|K_g)$ so that every valid response contributes equal
total weight, preventing longer query lists from dominating the local baseline.
The constants $\varepsilon_R$ and $\varepsilon_\ell$ stabilize the normalization of both signals.

Letting $\mathcal I_{gk}$ denote the token positions corresponding to the text of
$q_{gk}$, we combine the list-level and query-level advantages as
\begin{equation}
\widetilde A_{g,t}
=
A_g^{\mathrm{list}}
+
\lambda
\sum_{k=1}^{K_g}
\mathbf 1[t\in\mathcal I_{gk}]
A_{gk}^{\mathrm{loc}},
\label{eq:token_fusion}
\end{equation}
where $\lambda$ controls the local contribution.
All response tokens receive the list-level advantage, whereas query-text tokens
additionally receive their corresponding local advantage. Reasoning, priority labels,
and structural tokens retain only the list-level signal, and invalid responses receive
no local contribution. This assignment preserves overall list optimization while
allowing queries within the same response to receive distinct learning signals.

\paragraph{Policy Optimization.}
We optimize the policy using the two-level advantages, enabling query-specific updates
within the shared list-level objective while KL regularization constrains deviation
from the frozen SFT reference policy.
Writing $h_{g,t}=(\widetilde{s}_u,a_{<t}^{(g)})$, the probability ratio between the
current and rollout policies is
\begin{equation}
\rho_{g,t}(\theta)
=
\frac{
\pi_\theta(a_t^{(g)}\mid h_{g,t})
}{
\pi_{\mathrm{old}}(a_t^{(g)}\mid h_{g,t})
}.
\label{eq:policy_ratio}
\end{equation}
The policy is optimized using the following training loss:
\begin{equation}
\mathcal L_{\mathrm{RL}}(\theta)
=\mathbb E\!\left[\frac{1}{G}\sum_{g=1}^{G}\frac{1}{L_g}\sum_{t=1}^{L_g}
\bigl(-f_{\mathrm{dual}}(\rho_{g,t},\widetilde A_{g,t})+\beta\widehat D_{g,t}\bigr)\right],
\label{eq:rl_objective}
\end{equation}
where $L_g$ is the response length, $f_{\mathrm{dual}}$ is the dual-clipped surrogate,
and $\widehat D_{g,t}$ is the sampled KL penalty relative to the frozen SFT reference
policy $\pi_{\mathrm{ref}}$, weighted by $\beta$. The expectation is taken over training
users and sampled responses. The rollout policy is refreshed from the updated policy,
while rewards and advantages remain fixed during each update, as detailed in
Appendix~\ref{sec:grpo_details}.

By coordinating query relevance, service coverage, and response length, GRPO turns
target-domain behavioral feedback into targeted improvements in personalized service
query generation.

\section{Experiments}
\label{sec:experiments}
\subsection{Experimental Setup}
\label{sec:datasets}

\paragraph{Datasets.}
We construct \benchmark\ from real user profiles and heterogeneous map
interactions on a large-scale mapping platform. Following
Section~\ref{sec:data}, we obtain 19,006 user-level reasoning demonstrations
for SFT, while RL uses 3,000 separate users with subsequent target-domain
interaction feedback and another 500 users for validation. We evaluate on
user-disjoint \emph{MapRec-Rich} and \emph{MapRec-Sparse}, each containing
1,000 users across the same ten cities with rich and naturally sparse
mobility histories, respectively. Subsequent target-domain interactions
are held out as evaluation targets. Figure~\ref{fig:overview} illustrates
the service taxonomy and history distributions,
Table~\ref{tab:datasets} summarizes the dataset statistics, and
Appendix~\ref{sec:benchmark_details} provides further construction details.

\begin{table}[!tb]
\centering
\caption{Statistics of \benchmark. Per-user counts are means. The last row counts SFT queries or user--POI target pairs.}
\label{tab:datasets}
\small
\setlength{\tabcolsep}{3pt}
\renewcommand{\arraystretch}{1.10}
\begin{tabular*}{\columnwidth}{@{\extracolsep{\fill}}lrrrr@{}}
\toprule
& \multicolumn{2}{c}{Training}
& \multicolumn{2}{c}{Evaluation} \\
\cmidrule(lr){2-3}\cmidrule(lr){4-5}
Statistic & SFT & RL & Rich & Sparse \\
\midrule
Users              & 19,006    & 3,000   & 1,000   & 1,000 \\
History events     & 4,383,201 & 704,164 & 135,793 & 5,826 \\
Events/user        & 230.62    & 234.72  & 135.79  & 5.83 \\
Action types/user  & 5.33      & 5.37    & 3.07    & 1.63 \\
Profile fields/user & 6.65     & 7.19    & 6.63    & 5.65 \\
\midrule
Queries/targets    & 64,378    & 43,920  & 11,335  & 2,876 \\
\bottomrule
\end{tabular*}
\end{table}

\begin{table*}[!t]
\centering
\caption{Query-generation results on MapRec-Rich and MapRec-Sparse (\%).
POI retrieval uses spatial hybrid retrieval over a shared pool of
2 million POIs. All 1,000 users and all valid queries per benchmark
are evaluated. Bold marks column maxima.}
\label{tab:offline_main}
\small
\setlength{\tabcolsep}{2.8pt}
\renewcommand{\arraystretch}{1.08}
\setlength{\aboverulesep}{1pt}
\setlength{\belowrulesep}{1.5pt}
\begin{tabular*}{\textwidth}{@{\extracolsep{\fill}}lrrrrr@{\hspace{10pt}}rrrrr@{}}
\toprule
& \multicolumn{5}{c}{MapRec-Rich}
& \multicolumn{5}{c}{MapRec-Sparse} \\
\cmidrule(lr){2-6}\cmidrule(lr){7-11}
& \multicolumn{3}{c}{Retrieval utility} & Service & Semantic
& \multicolumn{3}{c}{Retrieval utility} & Service & Semantic \\
\cmidrule(lr){2-4}\cmidrule(lr){5-5}\cmidrule(lr){6-6}
\cmidrule(lr){7-9}\cmidrule(lr){10-10}\cmidrule(lr){11-11}
Model / Stage
& Hit@50 & Recall@50 & NDCG@50 & SIA@10 & SDA
& Hit@50 & Recall@50 & NDCG@50 & SIA@10 & SDA \\

\midrule
\rowcolor[gray]{0.94}
\multicolumn{11}{c}{%
\raisebox{-0.3ex}{\includegraphics[height=1.8ex]{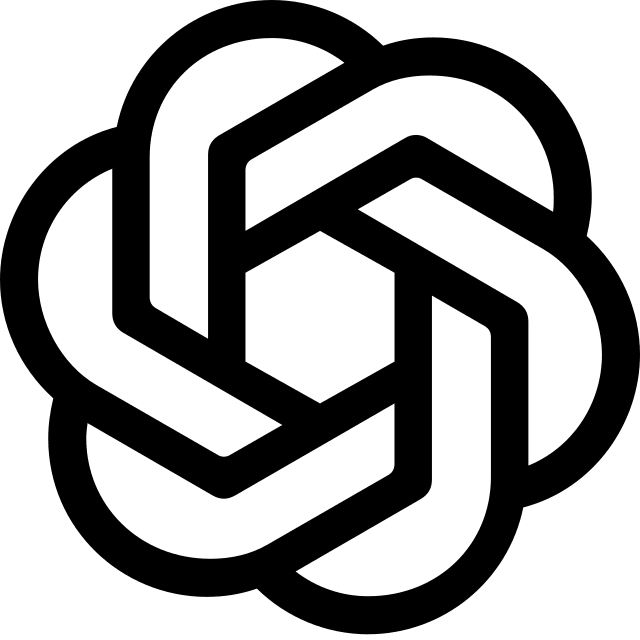}}\hspace{0.35em}%
\raisebox{-0.3ex}{\includegraphics[height=1.8ex]{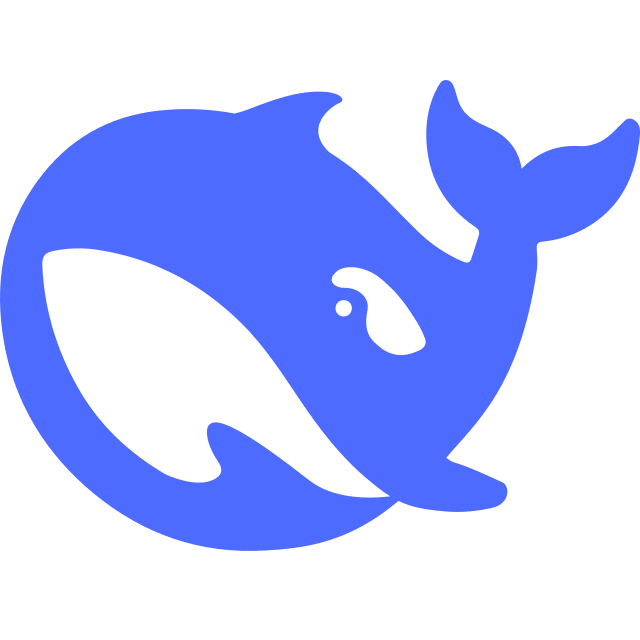}}\hspace{0.35em}%
\raisebox{-0.3ex}{\includegraphics[height=1.8ex]{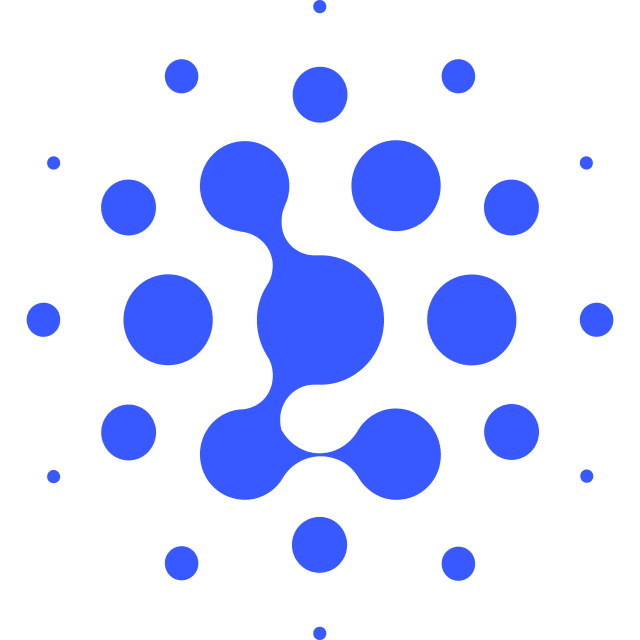}}\hspace{0.35em}%
\raisebox{-0.3ex}{\includegraphics[height=1.8ex]{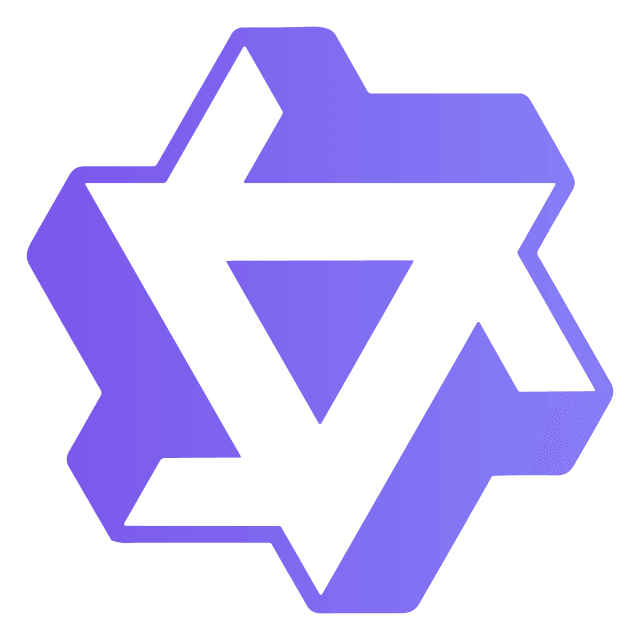}}\hspace{0.35em}%
\raisebox{-0.3ex}{\includegraphics[height=1.8ex]{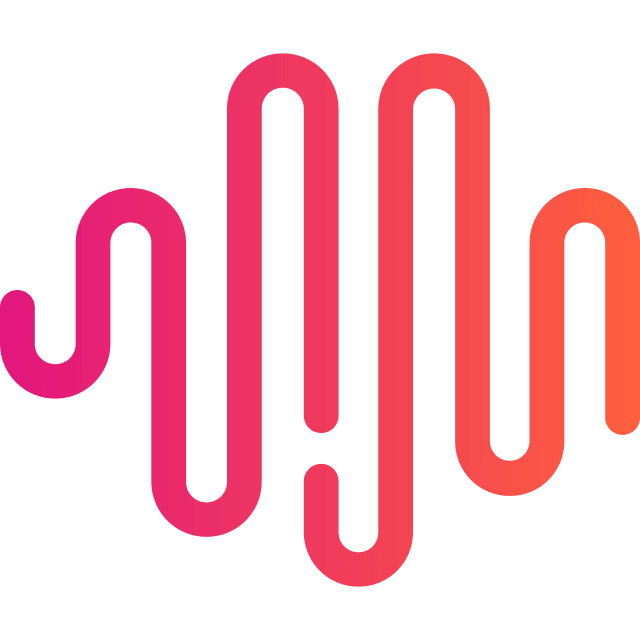}}\hspace{0.65em}%
\textbf{\textit{API-evaluated models}}} \\
\midrule
GPT-4o
& 1.70 & 0.164 & 0.178 & 31.19 & 35.28
& 0.50 & 0.197 & 0.109 & 13.78 & 15.02 \\
DeepSeek-v4-pro
& 1.30 & 0.157 & 0.158 & 28.94 & 30.89
& 0.50 & 0.277 & 0.201 & 12.46 & 12.82 \\
GLM-5.2
& 1.50 & 0.165 & 0.150 & 29.85 & 31.75
& 0.50 & 0.274 & 0.197 & 13.31 & 14.42 \\
Qwen3.8-Max
& 1.70 & 0.206 & 0.205 & 32.95 & 36.08
& 0.70 & 0.308 & 0.178 & 14.43 & 15.45 \\
MiniMax-M2.7
& 1.30 & 0.145 & 0.101 & 27.51 & 29.42
& 0.40 & 0.177 & 0.114 & 12.00 & 13.33 \\

\midrule
\rowcolor[gray]{0.94}
\multicolumn{11}{c}{%
\raisebox{-0.3ex}{\includegraphics[height=1.8ex]{figures/model-logos/openai.png}}\hspace{0.35em}%
\raisebox{-0.3ex}{\includegraphics[height=1.8ex]{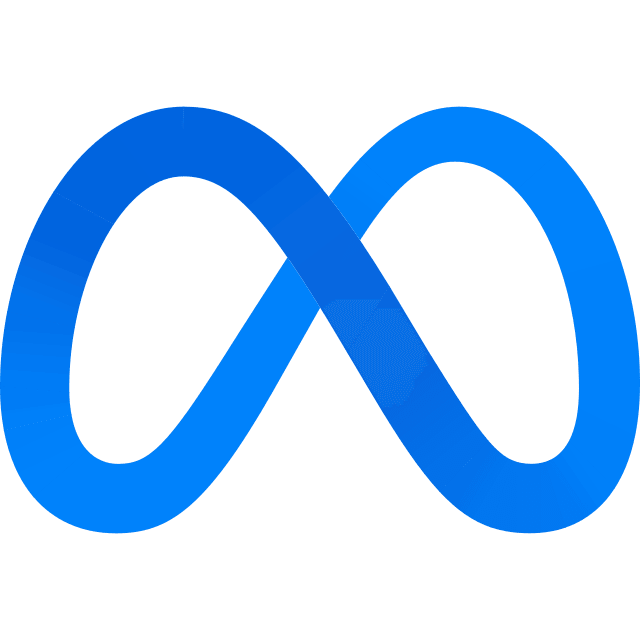}}\hspace{0.35em}%
\raisebox{-0.3ex}{\includegraphics[height=1.8ex]{figures/model-logos/qwen-logo.png}}\hspace{0.65em}%
\textbf{\textit{Locally evaluated models}}} \\
\midrule
GPT-OSS-120B
& 3.00 & 0.350 & \textbf{0.329} & 25.53 & 22.59
& \textbf{0.80} & 0.337 & 0.254 & 10.84 & 12.15 \\
Llama3.1-8B-Instruct
& 0.60 & 0.058 & 0.057 & 12.17 & 13.39
& 0.00 & 0.000 & 0.000 & 2.39 & 2.80 \\
Llama3.1-70B-Instruct
& 0.90 & 0.156 & 0.095 & 28.48 & 37.39
& 0.70 & 0.261 & 0.182 & 12.04 & 14.06 \\
Qwen3-235B-A22B
& 2.10 & 0.219 & 0.198 & 34.52 & 30.48
& 0.70 & 0.317 & 0.281 & 14.25 & 13.10 \\
Qwen3.5-397B-A17B
& 0.90 & 0.110 & 0.079 & 30.30 & 33.52
& 0.40 & 0.063 & 0.079 & 12.62 & 15.22 \\

\midrule
\rowcolor[gray]{0.94}
\multicolumn{11}{c}{
\raisebox{-0.3ex}{\includegraphics[height=1.8ex]{figures/model-logos/qwen-logo.png}}
\hspace{0.45em}\textbf{\textit{Backbone: Qwen3.5-4B}}} \\
\midrule
Base
& 1.30 & 0.194 & 0.131 & 21.35 & 20.99
& 0.40 & 0.192 & 0.108 & 8.90 & 9.55 \\
+ SFT
& 1.50 & 0.159 & 0.171 & 32.50 & 34.97
& 0.50 & 0.269 & 0.276 & 14.41 & 15.77 \\
\textbf{+ GRPO}
& \textbf{3.40} & 0.338 & 0.325 & 39.93 & 51.37
& \textbf{0.80} & \textbf{0.342} & \textbf{0.282} & 18.20 & 25.80 \\

\midrule
\rowcolor[gray]{0.94}
\multicolumn{11}{c}{
\raisebox{-0.3ex}{\includegraphics[height=1.8ex]{figures/model-logos/qwen-logo.png}}
\hspace{0.45em}\textbf{\textit{Backbone: Qwen3.5-9B}}} \\
\midrule
Base
& 0.90 & 0.134 & 0.102 & 24.21 & 27.26
& 0.30 & 0.146 & 0.143 & 11.05 & 12.60 \\
+ SFT
& 1.30 & 0.134 & 0.143 & 33.38 & 35.61
& 0.40 & 0.083 & 0.085 & 14.02 & 15.03 \\
\textbf{+ GRPO}
& 2.60 & 0.273 & 0.252 & 40.01 & 52.60
& 0.40 & 0.177 & 0.195 & 18.42 & 27.67 \\

\midrule
\rowcolor[gray]{0.94}
\multicolumn{11}{c}{
\raisebox{-0.3ex}{\includegraphics[height=1.8ex]{figures/model-logos/qwen-logo.png}}
\hspace{0.45em}\textbf{\textit{Backbone: Qwen3.5-27B}}} \\
\midrule
Base
& 1.50 & 0.166 & 0.160 & 31.79 & 31.53
& 0.50 & 0.300 & 0.131 & 12.69 & 14.70 \\
+ SFT
& 1.70 & 0.168 & 0.178 & 33.28 & 33.87
& 0.50 & 0.277 & 0.203 & 14.01 & 15.57 \\
\textbf{+ GRPO}
& 3.30 & \textbf{0.381} & 0.297 & \textbf{42.45} & \textbf{72.19}
& 0.60 & 0.293 & 0.213 & \textbf{21.35} & \textbf{43.43} \\

\midrule
\rowcolor[gray]{0.94}
\multicolumn{11}{c}{
\raisebox{-0.3ex}{\includegraphics[height=1.8ex]{figures/model-logos/qwen-logo.png}}
\hspace{0.45em}\textbf{\textit{Backbone: Qwen3.5-122B-A10B}}} \\
\midrule
Base
& 1.40 & 0.205 & 0.190 & 28.67 & 29.93
& 0.50 & 0.274 & 0.201 & 12.29 & 14.07 \\
+ SFT
& 2.00 & 0.222 & 0.211 & 32.81 & 33.79
& 0.60 & 0.203 & 0.141 & 14.07 & 14.51 \\
\textbf{+ GRPO}
& 3.10 & 0.331 & 0.258 & 41.32 & 61.51
& 0.70 & 0.295 & 0.192 & 19.63 & 34.35 \\

\bottomrule
\end{tabular*}
\end{table*}

\paragraph{Baselines.}
Table~\ref{tab:offline_main} compares general-purpose LLMs with Base, SFT,
and GRPO variants of Qwen3.5 at the 4B, 9B, 27B, and 122B-A10B scales.
For each backbone, + SFT and + GRPO denote successive training stages,
with + GRPO representing the complete \method\ model.
Table~\ref{tab:recommendation_baselines} adds cross-domain adaptations of
GLoSS~\cite{gloss}, Rec-R1~\cite{rec_r1}, and PIGEON~\cite{pigeon} built on
Qwen3.5-4B, where PIGEON-history uses historical service and context proxies
to generate a single query per user. We also evaluate Content-ComiRec-SA,
a content-based adaptation of ComiRec-SA~\cite{comirec} that replaces
item-ID embeddings with projected POI content embeddings to rank POIs
directly by multi-interest matching scores, while SIA and SDA apply only
to query-generating methods. Finally, Table~\ref{tab:grpo_components}
includes task-adapted Rank-GRPO~\cite{rank_grpo} as a policy-optimization
baseline.

\paragraph{Evaluation Metrics.}
We evaluate performance across three complementary dimensions. For
\emph{POI retrieval utility}, we report Hit@50, Recall@50, and
NDCG@50~\cite{ndcg} over a shared two-million-POI pool while retaining all
future targets, including those outside the pool. Query-based methods
share a spatial hybrid retrieval backend in which semantic and keyword
ranks are fused and geographically weighted, with each POI receiving its
maximum score across valid queries, whereas Content-ComiRec-SA retains
its native multi-interest ranking. For \emph{service intent alignment},
SIA@10 measures category-level NDCG@10 over 263 service categories ranked
by a frozen service matcher, using categories observed in future
interactions as relevance labels. For \emph{semantic demand alignment},
a frozen GPT-OSS-120B evaluator assesses core service matching, constraint
compatibility, specific-service matching, and target coverage, aggregated as
\begin{equation}
\mathrm{SDA}=\frac{1}{4N}\sum_{u=1}^{N}
(\mathrm{Core}_u+\mathrm{Compat}_u+\mathrm{Spec}_u+\mathrm{Cov}_u),
\label{eq:eval_sda}
\end{equation}
where $N$ is the number of test users. The first three components assess
individual queries, while coverage evaluates the complete query list
against future targets. All metrics are macro-averaged over all 1,000
users in each test set and reported as percentages, with full
definitions and evaluation details in
Appendix~\ref{sec:metric_details}.

\paragraph{Implementation Details.}
We train \method\ across 64 NVIDIA H20 GPUs using
LlamaFactory~\cite{llamafactory} for SFT and verl~\cite{hybridflow} for
GRPO, with detailed training configurations and hyperparameters provided
in Appendix~\ref{sec:grpo_details}.

\subsection{Main Results}
Table~\ref{tab:offline_main} compares \method\ with general-purpose LLMs
under a shared spatial hybrid retrieval backend over two million POIs.

\paragraph{Overall performance.}
\method\ combines strong target-domain POI retrieval with superior
service and semantic alignment. Its 27B variant achieves the highest
SIA@10 and SDA on both benchmarks alongside the highest Recall@50 on
MapRec-Rich. Meanwhile, the 4B variant achieves the highest Hit@50 on
MapRec-Rich and Recall@50 on MapRec-Sparse, while maintaining NDCG@50
competitive with the strongest general-purpose baselines. These results
demonstrate that intent-guided query generation effectively bridges
heterogeneous mobility behaviors with both concrete POI retrieval and
the local-service demands reflected in subsequent target-domain
interactions.

\paragraph{Effectiveness of two-stage training.}
{\emergencystretch=1em
Supervised fine-tuning initializes structured cross-domain intent
reasoning, whereas GRPO further refines its recommendation utility using
subsequent target-domain behavioral feedback. Across all four backbone
scales and both benchmarks, GRPO consistently improves Recall@50,
NDCG@50, SIA@10, and SDA over the corresponding SFT models, while Hit@50
improves in seven settings and remains tied in one. For the 4B backbone
on MapRec-Rich, Hit@50 more than doubles from 1.50\% to 3.40\%, and
NDCG@50 increases from 0.171\% to 0.325\%. These consistent gains support
reinforcement learning with target-domain feedback for aligning generated
intents with the service needs reflected in subsequent user interactions.
\par}

\paragraph{Performance across model scales.}
Peak performance across metrics is distributed across model sizes, as
the 4B variant leads several POI retrieval metrics while the 27B variant
delivers the strongest service and semantic alignment. Scaling up
backbone parameters alone therefore does not uniformly improve
cross-domain recommendation quality. Instead, the substantial gains
from GRPO across all four scales confirm the need to adapt general
language capabilities to mobility-grounded intent inference and query
generation.

\begin{table}[!t]
\centering
\caption{Recommendation baselines (\%). Query-based methods use Qwen3.5-4B; \textemdash\ denotes inapplicable query metrics.}
\label{tab:recommendation_baselines}
\small
\setlength{\tabcolsep}{1.5pt}
\renewcommand{\arraystretch}{1.06}
\begin{tabular*}{\columnwidth}{@{\extracolsep{\fill}}lrrrrr@{}}
\toprule
Method & \shortstack{Hit\\@50} & \shortstack{Recall\\@50}
& \shortstack{NDCG\\@50} & \shortstack{SIA\\@10} & SDA \\
\midrule
\multicolumn{6}{c}{\textit{MapRec-Rich}} \\
\midrule
SFT & 1.50 & 0.159 & 0.171 & 32.50 & 34.97 \\
GLoSS & 1.90 & 0.218 & 0.211 & 28.03 & 17.23 \\
Rec-R1 & 1.70 & 0.167 & 0.121 & 25.61 & 50.40 \\
PIGEON-history & 0.30 & 0.049 & 0.044 & 21.56 & 44.94 \\
Content-ComiRec-SA & 1.80 & 0.202 & 0.212 & \textemdash & \textemdash \\
\textbf{\method} & \textbf{3.40} & \textbf{0.338} & \textbf{0.325}
& \textbf{39.93} & \textbf{51.37} \\
\midrule
\multicolumn{6}{c}{\textit{MapRec-Sparse}} \\
\midrule
SFT & 0.50 & 0.269 & 0.276 & 14.41 & 15.77 \\
GLoSS & 0.30 & 0.157 & 0.091 & 13.91 & 6.61 \\
Rec-R1 & 0.60 & 0.318 & 0.266 & 12.27 & 24.85 \\
PIGEON-history & 0.00 & 0.000 & 0.000 & 12.56 & 22.57 \\
Content-ComiRec-SA & 0.70 & 0.226 & 0.183 & \textemdash & \textemdash \\
\textbf{\method} & \textbf{0.80} & \textbf{0.342} & \textbf{0.282}
& \textbf{18.20} & \textbf{25.80} \\
\bottomrule
\end{tabular*}
\end{table}


\paragraph{Comparison with recommendation methods.}
Table~\ref{tab:recommendation_baselines} compares \method\ with
domain-adapted query-based and direct multi-interest recommendation
baselines. \method\ achieves the best performance across all three
retrieval metrics on both benchmarks and leads both alignment metrics
among query-generating methods. Specifically, its NDCG@50 reaches
0.325\% on MapRec-Rich and 0.282\% on MapRec-Sparse, substantially
outperforming Content-ComiRec-SA at 0.212\% and 0.183\%, respectively.
Although PIGEON-history attains SDA scores of 44.94\% on MapRec-Rich and
22.57\% on MapRec-Sparse, it yields markedly lower POI retrieval
accuracy, indicating that service-level semantic alignment alone is
insufficient to recover specific target POIs. Together with the
consistent improvements over GLoSS and Rec-R1, these results validate
our cross-domain intent reasoning and two-level credit assignment.

\subsection{Ablation Studies}
\label{sec:ablation_studies}

We use Qwen3.5-4B to examine input evidence, supervision formats, and GRPO
components under the main evaluation protocol.

\begin{table}[!htbp]
\centering
\caption{SFT input ablations with Qwen3.5-4B (\%).
Bold marks column maxima.}
\label{tab:sft_input}
\small
\setlength{\tabcolsep}{1.5pt}
\renewcommand{\arraystretch}{1.08}
\begin{tabular*}{\columnwidth}{@{\extracolsep{\fill}}lrrr@{\hspace{5pt}}rrr@{}}
\toprule
& \multicolumn{3}{c}{MapRec-Rich}
& \multicolumn{3}{c}{MapRec-Sparse} \\
\cmidrule(lr){2-4}\cmidrule(lr){5-7}
Input
& \shortstack{NDCG\\@50} & \shortstack{SIA\\@10} & SDA
& \shortstack{NDCG\\@50} & \shortstack{SIA\\@10} & SDA \\
\midrule
Full input
& \textbf{0.171} & \textbf{32.50} & 34.97
& \textbf{0.276} & \textbf{14.41} & \textbf{15.77} \\
No profile
& 0.168 & 31.80 & \textbf{35.51}
& 0.185 & 12.93 & 13.35 \\
No history
& 0.000 & 15.24 & 19.15
& 0.000 & 7.77 & 8.32 \\
7-day history
& 0.041 & 23.90 & 26.61
& 0.147 & 5.90 & 5.63 \\
\bottomrule
\end{tabular*}
\end{table}

\paragraph{Input evidence.}
Personalized intent inference connects user characteristics with
heterogeneous mobility evidence. Table~\ref{tab:sft_input} compares
independently trained models with the same input intervention during
SFT and evaluation. All variants retain the original teacher targets.
Removing interaction history reduces NDCG@50 to
0.000\% on both benchmarks while substantially lowering service and
semantic alignment, underscoring the importance of mobility context for
retrieving relevant target-domain POIs. Removing user profiles yields
mixed results on MapRec-Rich but lowers all three metrics on
MapRec-Sparse, showing that profile attributes provide complementary
personalization when behavioral evidence is sparse. Retaining the full
sixty-day history also outperforms the seven-day variant across all
three metrics on both benchmarks, supporting the use of broader
temporal mobility context.

\begin{table}[!htbp]
\centering
\caption{SFT supervision ablations with Qwen3.5-4B (\%).
Bold marks column maxima.}
\label{tab:sft_supervision}
\small
\setlength{\tabcolsep}{1.5pt}
\renewcommand{\arraystretch}{1.08}
\begin{tabular*}{\columnwidth}{@{\extracolsep{\fill}}lrrr@{\hspace{5pt}}rrr@{}}
\toprule
& \multicolumn{3}{c}{MapRec-Rich}
& \multicolumn{3}{c}{MapRec-Sparse} \\
\cmidrule(lr){2-4}\cmidrule(lr){5-7}
Supervision
& \shortstack{NDCG\\@50} & \shortstack{SIA\\@10} & SDA
& \shortstack{NDCG\\@50} & \shortstack{SIA\\@10} & SDA \\
\midrule
Full SFT
& \textbf{0.171} & \textbf{32.50} & \textbf{34.97}
& \textbf{0.276} & \textbf{14.41} & 15.77 \\
Direct-SFT
& 0.137 & 31.69 & 31.41
& 0.191 & 14.35 & \textbf{15.95} \\
Analysis-SFT
& 0.128 & 30.33 & 32.70
& 0.204 & 14.14 & 14.55 \\
\bottomrule
\end{tabular*}
\end{table}

\paragraph{SFT supervision.}
Structured supervision connects raw mobility evidence, intermediate
intent reasoning, and prioritized query generation.
Table~\ref{tab:sft_supervision} compares three supervision formats
trained on the same 19,006 users for the same number of optimization
steps. Full SFT retains evidence restatement, three-stage intent
analysis, and prioritized queries, whereas Direct-SFT retains only
queries and Analysis-SFT removes evidence restatement. Full SFT achieves
the highest NDCG@50 and SIA@10 on both benchmarks and the highest SDA
on MapRec-Rich, while Direct-SFT attains a slightly higher SDA on
MapRec-Sparse. These results support retaining both evidence grounding
and intermediate reasoning to connect mobility understanding
with retrieval-oriented query generation, establishing a task-adapted
policy for reinforcement learning.

\begin{table}[!htbp]
\centering
\caption{GRPO variants and component ablations with Qwen3.5-4B (\%).
Bold marks column maxima.}
\label{tab:grpo_components}
\small
\setlength{\tabcolsep}{1.2pt}
\renewcommand{\arraystretch}{1.08}
\begin{tabular*}{\columnwidth}{
@{\extracolsep{\fill}}lcrrr@{\hspace{5pt}}rrr@{}
}
\toprule
 & & \multicolumn{3}{c}{MapRec-Rich}
   & \multicolumn{3}{c}{MapRec-Sparse} \\
\cmidrule(lr){3-5}\cmidrule(lr){6-8}
Variant & $\lambda$
& \shortstack{NDCG\\@50} & \shortstack{SIA\\@10} & SDA
& \shortstack{NDCG\\@50} & \shortstack{SIA\\@10} & SDA \\
\midrule
Rank-GRPO & \textemdash
& 0.098 & 33.55 & \textbf{61.45}
& 0.147 & 17.82 & \textbf{35.29} \\
\midrule
List only & 0
& 0.317 & 37.31 & 37.96
& 0.217 & 16.68 & 20.16 \\

Shared credit & 0.5
& 0.295 & 37.59 & 41.04
& 0.234 & 15.68 & 21.37 \\

Full method & 0.5
& 0.325 & \textbf{39.93} & 51.37
& 0.282 & \textbf{18.20} & 25.80 \\

Relaxed & 0
& 0.262 & 37.57 & 42.08
& \textbf{0.288} & 16.45 & 20.94 \\

Relaxed & 0.5
& \textbf{0.338} & \textbf{39.93} & 52.18
& 0.208 & 17.51 & 24.49 \\
\bottomrule
\end{tabular*}
\end{table}

\paragraph{GRPO components.}
Table~\ref{tab:grpo_components} examines credit allocation and coverage
matching in GRPO, with $\lambda$ controlling local credit in our internal
ablations. Under one-to-one matching, the full method improves
all three metrics on both benchmarks over the list-only variant with
$\lambda=0$ and the shared-credit variant. Shared credit preserves the
token-weighted sum of local advantages by broadcasting their
token-weighted mean across all query-text tokens. This comparison
supports assigning local feedback to the corresponding query under
shared list-level coverage guidance. A context-blinded evaluation also
yields higher context-supported coverage estimates for Full than Shared
with complete outputs under both LLM judges
(Table~\ref{tab:context_credit}). Compared with task-adapted
Rank-GRPO~\cite{rank_grpo}, the full method achieves higher NDCG@50 and
SIA@10 on both benchmarks despite lower SDA, showing that stronger
semantic assessment scores do not necessarily translate into better
retrieval of specific target POIs. With query-specific local credit,
replacing one-to-one matching with relaxed matching slightly improves
NDCG@50 and SDA on MapRec-Rich but lowers all three metrics on
MapRec-Sparse. Among internal ablations on MapRec-Sparse, the full
method leads SIA@10 and SDA, whereas relaxed matching without local
credit leads NDCG@50.

\subsection{Further Analysis}
\label{sec:further_analysis}

{\emergencystretch=1em
We analyze query budget, query expression, and human preference using
Qwen3.5-4B outputs under the main evaluation protocol.
\par}

\begin{figure}[tb]
\centering
\includegraphics[width=\columnwidth]{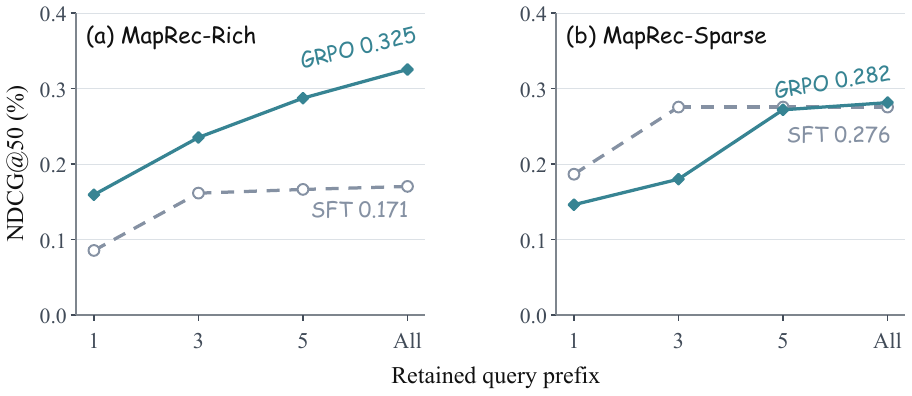}
\caption{POI retrieval across query budgets. All denotes the complete query list.}
\label{fig:query_budget}
\end{figure}

\paragraph{Query budget.}
Figure~\ref{fig:query_budget} compares prefixes of one, three, or five
queries against complete query lists. SFT gains largely saturate within
three queries, whereas GRPO continues to benefit from later queries. On
MapRec-Rich, GRPO leads at every budget, with NDCG@50 rising from 0.235\%
at three queries to 0.325\% with complete lists. On MapRec-Sparse, SFT
leads at one, three, and five queries, whereas GRPO surpasses SFT with
complete lists at 0.282\% versus 0.276\%, indicating greater retrieval
gains from complete lists.

\begin{figure}[tb]
\centering
\includegraphics[width=\columnwidth]{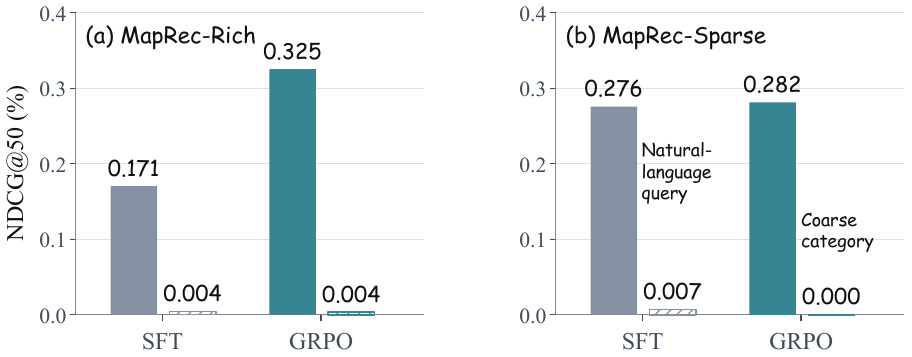}
\caption{POI retrieval with natural-language queries versus mapped coarse service category labels.}
\label{fig:query_expression}
\end{figure}

\paragraph{Query expression.}
Figure~\ref{fig:query_expression} replaces each generated query with its
top coarse service category predicted by the frozen matcher while
preserving query slots and the retrieval pipeline. NDCG@50 drops sharply
for both SFT and GRPO on both benchmarks, falling for GRPO from 0.325\%
to 0.004\% on MapRec-Rich and from 0.282\% to 0.000\% on MapRec-Sparse.
While category mapping also affects this comparison, the results support
retaining descriptive query content beyond coarse service labels.

\paragraph{Human assessment.}
In side-by-side human evaluation on MapRec-Sparse in
Table~\ref{tab:human_evaluation}, annotators consistently favor GRPO over
both Base and SFT with preference rates of 70\% to 80\%.

Appendix~\ref{sec:case_study} presents a qualitative case in
Figure~\ref{fig:case_study}.

\section{Conclusion}
{\emergencystretch=1em
We presented \method\ for intent-based cross-domain POI recommendation,
bridging heterogeneous mobility behaviors and user profiles with
personalized local service retrieval by combining progressive reasoning
supervision with intent-aligned GRPO. Leveraging subsequent behavioral
feedback from the target domain, our two-level credit assignment
mechanism jointly optimizes individual query relevance and collective
service coverage while suppressing redundancy. Extensive experiments
on \benchmark\ across four backbone scales demonstrate consistent gains
in both concrete POI retrieval utility and semantic demand alignment
across rich and sparse mobility histories, underscoring the value of
implicit intent inference in connecting everyday mobility traces with
downstream local service discovery.
\par}

\clearpage
\balance
\bibliographystyle{ACM-Reference-Format}
\bibliography{references}
\clearpage
\nobalance
\appendix
\section{MILES-Bench Details}
\label{sec:benchmark_details}
Sections~\ref{sec:problem} and~\ref{sec:data} define disjoint user splits, pre-prediction inputs, and future targets deduplicated by POI per user. RL validation contains 500 users and 7,507 user--POI pairs. MapRec-Rich includes 100 users each from Beijing, Shanghai, Guangzhou, Shenzhen, Hangzhou, Nanjing, Wuhan, Chengdu, Chongqing, and Xi'an. MapRec-Sparse includes 200 users per history-length band of 0, 1--2, 3--5, 6--10, or 11--20 events, retaining available profiles and city context. Figure~\ref{fig:overview} shows 263 fine categories within twelve broad groups. Table~\ref{tab:benchmark_characteristics} provides additional statistics.

\begin{table}[!htb]
\centering
\caption{Additional benchmark characteristics. Multi-service users have targets in at least two broad categories.}
\label{tab:benchmark_characteristics}
\small
\setlength{\tabcolsep}{2pt}
\renewcommand{\arraystretch}{1.08}
\begin{tabular*}{\columnwidth}{@{\extracolsep{\fill}}lrr@{}}
\toprule
Statistic & MapRec-Rich & MapRec-Sparse \\
\midrule
Median history events/user & 104.5 & 4.0 \\
90th-percentile history events & 265 & 15 \\
History-length range & 7--1,002 & 0--20 \\
Distinct target POIs & 10,872 & 2,832 \\
Observed fine categories & 208 & 164 \\
Mean broad categories/user & 3.95 & 1.71 \\
Multi-service users (\%) & 94.10 & 40.20 \\
History-unseen target pairs (\%) & 97.27 & 97.18 \\
\bottomrule
\end{tabular*}
\end{table}
\paragraph{Privacy protection.}
Mobility logs, vehicle attributes, and family information are de-identified before dataset construction, training, and evaluation. We will release a de-identified version of \benchmark\ and our code upon acceptance.

\section{Implementation Details}
\label{sec:grpo_details}
\subsection{Demonstrations and Service Matcher}
Qwen-235B generates three-stage demonstrations with 1--10 queries labeled \texttt{primary} or \texttt{expansion}, ordered primary-first. Checks cover required fields, nonempty analysis and queries, labels, duplicates, and event citations. GPT-5.1, DeepSeek-Pro, and human review assess evidence fidelity, temporal consistency, reasoning coherence, and intent distinctness. SFT supervises assistant tokens across 19,006 retained demonstrations containing user profile facts, cited events, staged reasoning, and prioritized queries.

Qwen-235B labels generated queries and historical searches paired with service definitions as \texttt{match}, \texttt{compatible}, or \texttt{mismatch}, excluding uncertain cases. User- and normalized-query-disjoint splits contain 621,949 training, 39,737 validation, and 38,545 test pairs. We fully fine-tune BGE-reranker-v2-m3~\cite{bge_reranker} with a three-class head using cross-entropy and AdamW for three epochs, with batch size 512, length 256, weight decay 0.01, 3\% warmup, and linear decay. Learning rates are $2\times10^{-5}$ for pretrained parameters and $10^{-4}$ for the head. The best validation macro-F1 checkpoint and Qwen3-Embedding-8B~\cite{qwen3_embedding} remain frozen during GRPO.

\subsection{Reward and Optimization Details}
SFT freezes the vision encoder and projector, training all language-model parameters with AdamW~\cite{adamw}, BF16, cosine decay, two epochs, batch size 64, and length 98,304. Learning rates are $10^{-5}$ for 4B and 9B, and $5\times10^{-6}$ for 27B and 122B-A10B. GRPO uses 125 updates at $5\times10^{-7}$, sampling 24 users and eight responses per user. We set temperature to 1.0, top-$p$ to 0.95, $\beta=0.02$, $\lambda=0.5$, and input and output limits to 49,152 and 8,192.

Targets are grouped by service-category path for one-to-one matching in Section~\ref{sec:rl}. Broad-only groups include all child categories and are omitted if that user has a fine-category target under the same parent. A later query is redundant if normalized texts match, character-bigram Jaccard similarity is $\geq0.8$, or cosine similarity is $\geq0.95$ with the same best service group and no score improvement. Redundant queries contribute zero coverage, incur a 0.05 list penalty, and receive local score $-0.05$. Total response length $L_g$ incurs

\begin{equation}
L_{\mathrm{len}}(a^{(g)})=0.05\,\operatorname{clip}\!\left(\frac{L_g-2{,}048}{6{,}144},0,1\right).
\label{eq:appendix_length_penalty}
\end{equation}
Invalid or truncated responses receive zero reward and no local correction. List advantages cover all response tokens, while local advantages cover corresponding query text. Local normalization weights valid responses equally, with both stabilizers $10^{-6}$. Equation~\eqref{eq:rl_objective} uses dual clipping with $\epsilon=0.2$ and $c=3$:

\begin{equation}
\begin{aligned}
f_{\mathrm{clip}}(\rho,A)&=\min\{\rho A,\operatorname{clip}(\rho,1-\epsilon,1+\epsilon)A\},\\
f_{\mathrm{dual}}(\rho,A)&=\begin{cases}
\max\{f_{\mathrm{clip}}(\rho,A),cA\},&A<0,\\
f_{\mathrm{clip}}(\rho,A),&A\geq0.
\end{cases}
\end{aligned}
\label{eq:dual_clip}
\end{equation}
The sampled KL term is
\begin{equation}
\widehat D_{g,t}=v_{g,t}-\log v_{g,t}-1,\qquad
v_{g,t}=\frac{\pi_{\mathrm{ref}}(a_t^{(g)}\mid h_{g,t})}{\pi_\theta(a_t^{(g)}\mid h_{g,t})}.
\label{eq:kl_estimate}
\end{equation}
The reference policy remains fixed at the SFT model, while rollouts follow the updated policy. Group rewards and token-level advantages remain fixed within each policy update.

\begin{figure}[t]
    \centering
    \includegraphics[width=\columnwidth]{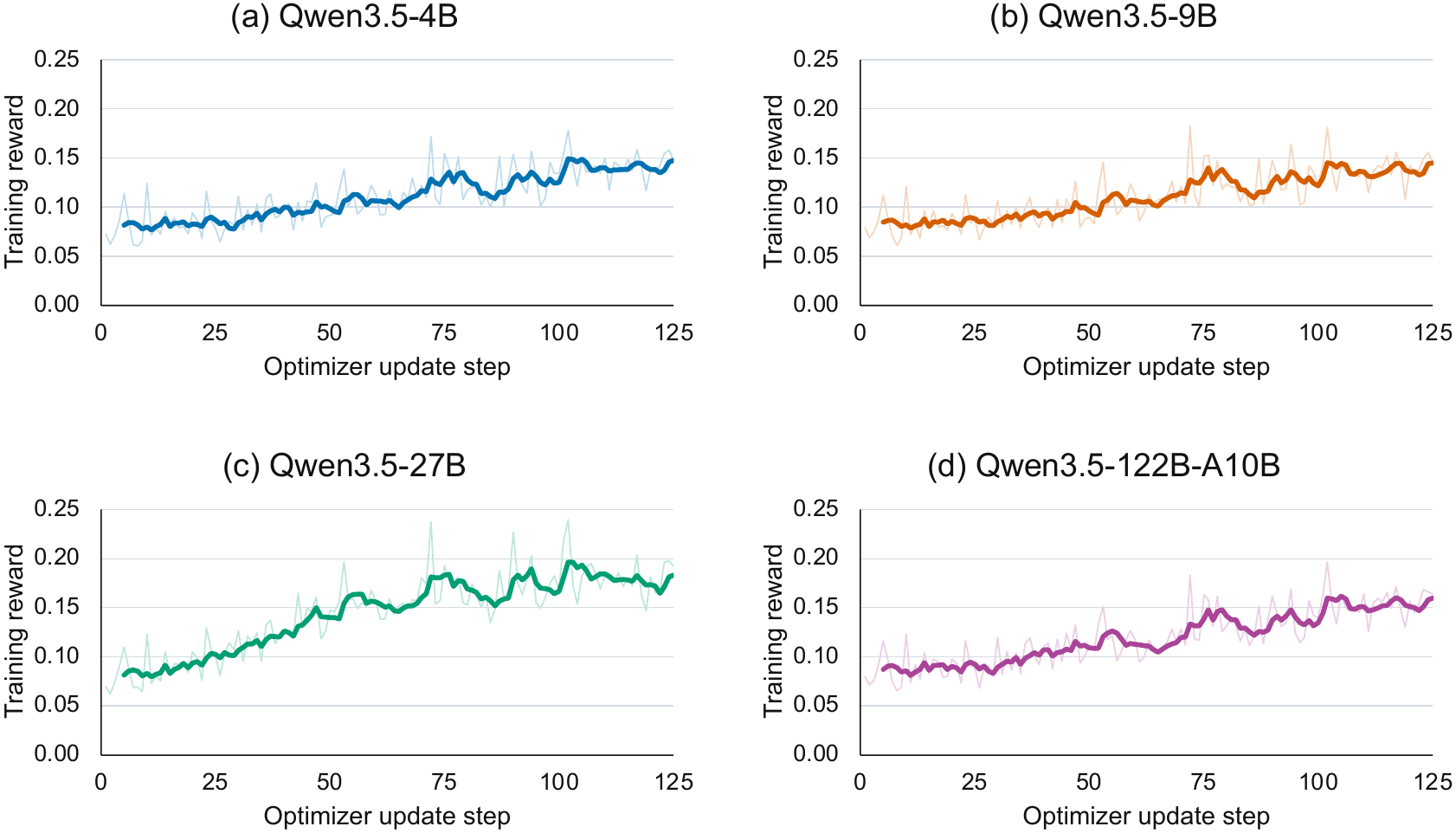}
    \caption{GRPO training rewards across four backbone sizes.}
    \label{fig:grpo_training_curves}
\end{figure}

\paragraph{GRPO training dynamics.}
Figure~\ref{fig:grpo_training_curves} shows overall upward trends in training rewards across all four backbones over 125 updates, with the 27B variant achieving the highest rewards at later steps.

\paragraph{Near-line deployment.}
We are developing asynchronous intent inference triggered by searches and navigation, caching queries and embeddings for online POI retrieval and ranking.

\section{Evaluation Details}
\label{sec:metric_details}
Metrics are macro-averaged over all 1,000 users and reported as percentages. Query-based methods share frozen retrieval and evaluation models. Content-ComiRec-SA uses native POI scores for retrieval utility only. Empty queries and later whitespace-normalized duplicates become empty positions. Retrieval uses all valid queries, while SDA retains original positions in its denominator. Outputs without valid queries score zero.

\subsection{POI Retrieval Utility}
A fixed seed samples two million POIs without replacement, proportional to retrievable catalog sizes across ten cities, without city prefiltering or target injection. Each query retrieves the top 100 with Qwen3-Embedding-8B and top 1,000 each with field-BM25~\cite{bm25} and term-boosted BM25. Dense scores are recomputed over the deduplicated union across queries. Scaled reciprocal rank fusion~\cite{rrf} combines dense and field-BM25 ranks:

\begin{equation}
H(q,p)=\frac{61}{2}\left(
\frac{1}{60+r_d}
+\frac{\mathbf{1}[p\in B_q]}{60+r_b}
\right),
\label{eq:hybrid_fusion}
\end{equation}
Here $r_d$ ranks the union and $r_b$ ranks the field-BM25 set $B_q$; term-boosted BM25 only expands candidates. Geographic weighting is $G(q,p)=0.85+0.15/(1+d(q,p)/\tau_q)$, with $d$ in kilometers. Uniquely resolved locations use $\tau_q=2$. Without explicit locations, $\tau_q=5$ and $d$ measures distance to the nearest of up to three historical anchors weighted by frequency and recency with a 14-day half-life. Unresolved locations or missing coordinates give $G=1$. User scoring is

\begin{equation}
s_u(p)=\max_{q\in \mathcal Q_u^{\mathrm{valid}}}H(q,p)G(q,p).
\label{eq:eval_ranking}
\end{equation}
Hit@50, Recall@50, and NDCG@50 are computed over distinct future target POIs. All ground-truth targets remain in metric denominators, including those outside the candidate pool.

\subsection{Service Intent Alignment}
Target categories are deduplicated within the 263-category catalog and ranked by the frozen matcher's maximum query score:

\begin{equation}
 s_u^{\mathrm E}(c)=\max_{q\in\mathcal Q_u^{\mathrm{valid}}}p_{\mathrm{match}}(q,c).
\label{eq:eval_service_score}
\end{equation}
SIA@10 is category-level Exact-NDCG@10. Unmapped categories are excluded; users without mapped targets score zero.

\subsection{Semantic Demand Alignment}
GPT-OSS-120B receives queries, target POI names, and business categories, excluding model identities, reasoning, and retrieval scores. We use temperature 0, top-$p$ 1, seed 42, and medium reasoning effort. Core service relations are specific, broad, unrelated, or unknown, weighted $w=1,0.5,0,0$, respectively. Brand, location, price, and facility constraints are labeled supported, contradicted, or unknown. Compatibility $a=1$ iff none is contradicted. Unknown constraints remain unverified. Pair scores are

\begin{equation}
\begin{aligned}
 f_{\mathrm{core}}(q,p)&=w(q,p),\\
 f_{\mathrm{compat}}(q,p)&=w(q,p)a(q,p),\\
 f_{\mathrm{spec}}(q,p)&=\mathbf1[w(q,p)=1]a(q,p).
\end{aligned}
\label{eq:eval_semantic_pair}
\end{equation}
These scores distinguish core relevance, constraint compatibility, and specific-service agreement without requiring the same store. For future target POIs $G_u$, original query count $m_u$, and $d_u=\max(1,m_u)$, matching and coverage are

\begin{equation}
\begin{aligned}
 \mathrm{Match}^{(v)}_u
 &=\frac1{d_u}\sum_{j=1}^{m_u}\max_{p\in G_u}f_v(q_{u,j},p),\\
 \mathrm{Cov}_u
 &=\frac{|\{p\in G_u:\exists j,\ f_{\mathrm{spec}}(q_{u,j},p)=1\}|}{|G_u|},
\end{aligned}
\label{eq:eval_judge}
\end{equation}
where $v\in\{\mathrm{core},\mathrm{compat},\mathrm{spec}\}$. Matching averages each query's best match, whereas coverage measures the list's matched target fraction. Empty and duplicate positions contribute zero. Each query can cover multiple targets, each counted once. Equation~\eqref{eq:eval_sda} averages all four components before rounding.

\section{Additional Results and Analysis}
\label{sec:offline_supplement}
\subsection{Semantic Components}
Table~\ref{tab:semantic_gains} shows that all four SDA components improve from SFT to GRPO across four backbones and both benchmarks.

\begin{table}[!htb]
\centering
\caption{SDA component gains from SFT to GRPO in percentage points, computed before rounding.}
\label{tab:semantic_gains}
\small
\setlength{\tabcolsep}{2pt}
\renewcommand{\arraystretch}{1.08}
\begin{tabular*}{\columnwidth}{@{\extracolsep{\fill}}lrrrr@{}}
\toprule
Backbone & Core & Compat. & Spec. & Cov. \\
\midrule
\multicolumn{5}{c}{\textit{MapRec-Rich}} \\
4B & +11.89 & +12.24 & +6.78 & +34.67 \\
9B & +11.87 & +12.93 & +7.60 & +35.57 \\
27B & +29.28 & +36.59 & +34.42 & +53.01 \\
122B-A10B & +20.64 & +25.15 & +21.44 & +43.64 \\
\midrule
\multicolumn{5}{c}{\textit{MapRec-Sparse}} \\
4B & +8.01 & +8.93 & +3.74 & +19.44 \\
9B & +9.55 & +10.51 & +4.93 & +25.57 \\
27B & +24.43 & +26.15 & +17.38 & +43.48 \\
122B-A10B & +15.85 & +17.60 & +11.73 & +34.15 \\
\bottomrule
\end{tabular*}
\end{table}

\paragraph{Human assessment of query quality.}
Two human assessors independently compare anonymous 4B GRPO outputs with Base and SFT on 100 MapRec-Sparse users. Table~\ref{tab:human_evaluation} reports preference rates of 70\% to 80\%, calculated as $(\mathrm{Win}+0.5\,\mathrm{Tie})/100$, supporting improved overall query quality.

\begin{table}[!htb]
\centering
\caption{Human preferences for 4B GRPO on MapRec-Sparse. Each row contains 100 comparisons.}
\label{tab:human_evaluation}
\small
\setlength{\tabcolsep}{2pt}
\renewcommand{\arraystretch}{1.05}
\begin{tabular*}{\columnwidth}{@{\extracolsep{\fill}}llrrrr@{}}
\toprule
Assessor & Comparator & Win & Tie & Loss & Pref. (\%) \\
\midrule
1 & 4B Base & 61 & 18 & 21 & 70.00 \\
1 & 4B SFT  & 72 & 16 & 12 & 80.00 \\
2 & 4B Base & 65 & 12 & 23 & 71.00 \\
2 & 4B SFT  & 64 & 18 & 18 & 73.00 \\
\bottomrule
\end{tabular*}
\end{table}

\paragraph{Context-supported service coverage.}
GPT-4o and DeepSeek-v4-pro independently judge anonymous 4B Full and Shared outputs for 100 users per benchmark. They assess intent and key-constraint support from profiles and histories before seeing future targets, then keep these judgments fixed. Coverage counts future categories matched by supported queries, excluding unknown or conflicting key constraints. Each category counts once in the 12-category taxonomy, with all future categories retained in the denominator. Sparse averages use sampling-design weights.

Table~\ref{tab:context_credit} shows that Full consistently achieves higher context-supported coverage than Shared credit under both LLM judges on MapRec-Rich and MapRec-Sparse, for both five-query prefixes and complete lists. These results support the effectiveness of query-specific credit assignment in improving grounded service coverage.

\begin{table}[!htb]
\centering
\caption{Context-supported coverage (\%). $\Delta$: Full minus Shared in percentage points. DeepSeek differences use displayed scores.}
\label{tab:context_credit}
\small
\setlength{\tabcolsep}{2pt}
\renewcommand{\arraystretch}{1.05}
\begin{tabular*}{\columnwidth}{@{\extracolsep{\fill}}llrrrr@{}}
\toprule
& & \multicolumn{3}{c}{DeepSeek-v4-pro} & GPT-4o \\
\cmidrule(lr){3-5}\cmidrule(lr){6-6}
Queries & Dataset & Full & Shared
& \multicolumn{1}{c}{$\Delta$}
& \multicolumn{1}{c@{}}{$\Delta$} \\
\midrule
First five & Rich   & 50.15 & 49.55 & $+0.60$ & $+0.69$ \\
           & Sparse & 32.71 & 31.97 & $+0.74$ & $+3.64$ \\
\midrule
All        & Rich   & 64.37 & 63.66 & $+0.71$ & $+0.63$ \\
           & Sparse & 42.71 & 39.12 & $+3.59$ & $+5.55$ \\
\bottomrule
\end{tabular*}
\end{table}

\subsection{Retrieval Analysis}
\label{sec:retrieval_analysis}

These analyses use saved 4B outputs under the main protocol.

\paragraph{Primary versus expansion queries.}
For users with both query types, $n_u^P$ and $n_u^E$ count primary and expansion queries. We compare the first $\min(n_u^P,n_u^E)$ of each type. Primary NDCG@50 is higher for both SFT and GRPO on both benchmarks. GRPO's primary and expansion scores are 0.1823\% and 0.1812\% on MapRec-Rich, and 0.1587\% and 0.1250\% on MapRec-Sparse, respectively.

\paragraph{Query ordering under a fixed budget.}
Using up to five GRPO queries, original, random, and reversed orders yield NDCG@50 of 0.2874\%, 0.2383\%, and 0.1779\% on Rich and 0.2721\%, 0.2404\%, and 0.2424\% on Sparse. Random ordering averages 20 permutations. Original order gains 0.0491 and 0.0317 percentage points over random, supporting prioritization under limited budgets.

\begin{figure}[!htb]
\centering
\includegraphics[width=\columnwidth]{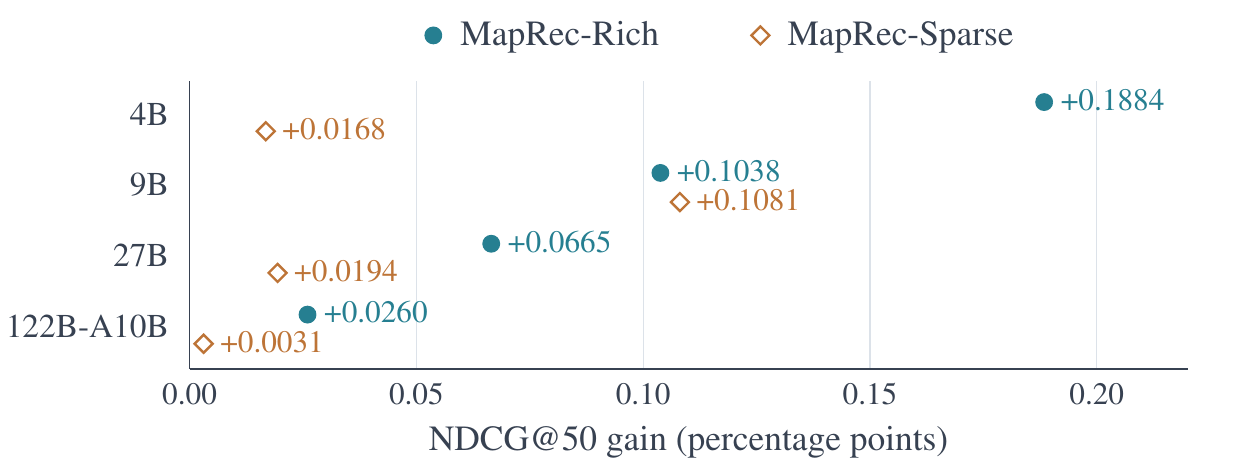}
\caption{NDCG@50 gains of GRPO over SFT with BGE-M3.}
\label{fig:bge_m3}
\end{figure}

\paragraph{Transfer across retrieval encoders.}
With the pool, lexical retrieval, and geographic weighting fixed, reward-independent BGE-M3~\cite{bge_m3} yields positive GRPO-over-SFT NDCG@50 gains of 0.0031--0.1884 percentage points across all scales and benchmarks in Figure~\ref{fig:bge_m3}.

\subsection{Case Study}
\label{sec:case_study}
\begin{figure}[H]
    \centering
    \includegraphics[width=0.75\columnwidth]{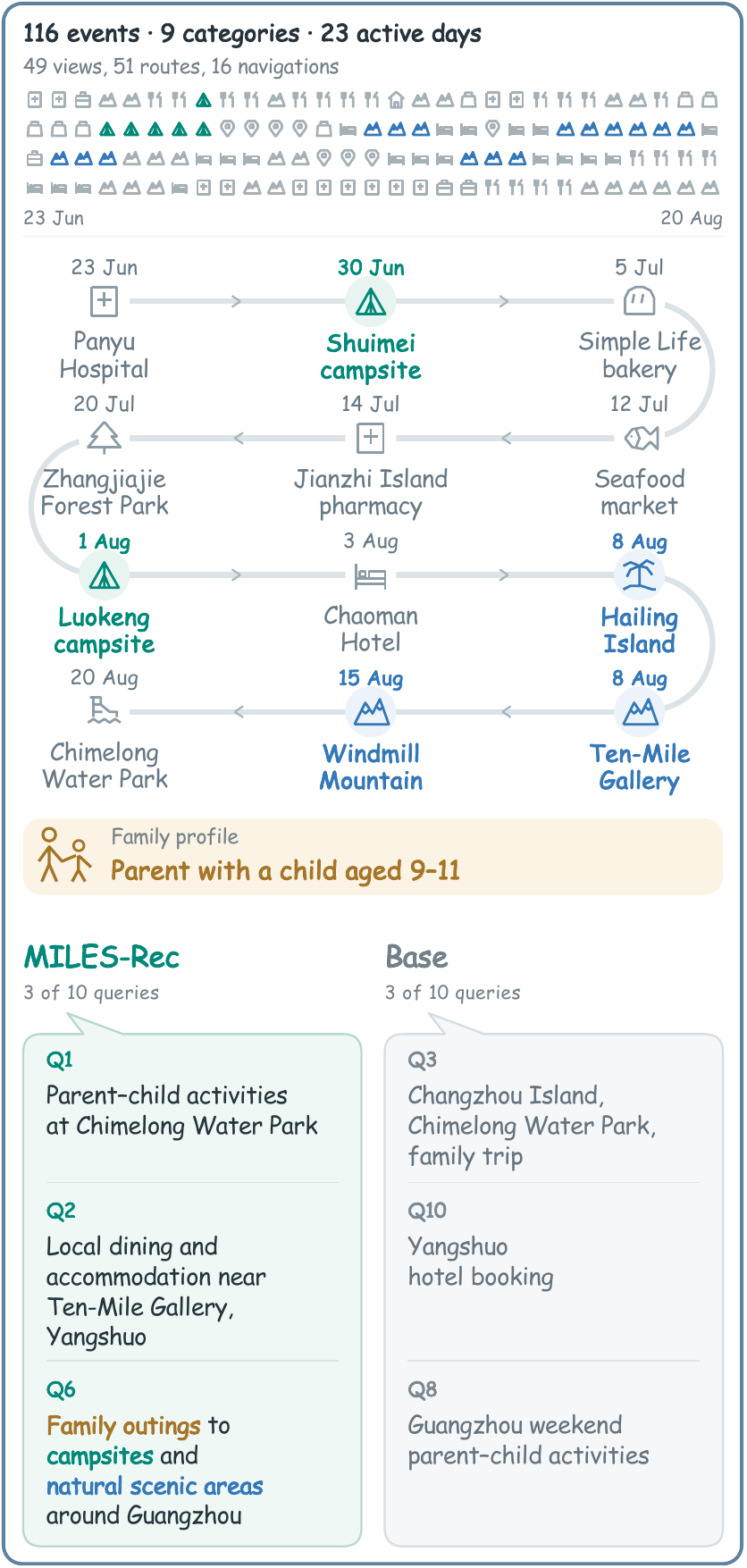}
    \caption{A qualitative comparison of MILES-Rec and Base. We show 12 of 116 historical interactions and three selected queries per model, retaining their original query indices.}
    \label{fig:case_study}
\end{figure}

Figure~\ref{fig:case_study} compares queries for a parent with campsite, scenic-area, and water-park histories. The Chimelong Water Park interaction and parent profile ground MILES-Rec's parent--child activities in Q1. The Ten-Mile Gallery interaction and earlier hotel and dining interests support Q2's complementary services. Repeated campsite and scenic-area interactions inform Q6's broader family outings, combining several observed interests.

\end{document}